\documentclass[prd,10pt,twocolumn,showpacs,preprintnumbers]{revtex4-2}
\usepackage{amsmath}

\usepackage{mathabx}
\usepackage{color} 
\usepackage{epsfig}
\usepackage{mathtools}
\usepackage{amsthm}
\usepackage{scalerel}

\newcounter{mnotecount}

\newcommand{\mnotex}[1]
{\protect{\stepcounter{mnotecount}}$^{\mbox{\footnotesize $\bullet$\themnotecount}}$ 
\marginpar{
\raggedright\tiny\em
$\!\!\!\!\!\!\,\bullet$\themnotecount: #1} }

\allowdisplaybreaks

\newtheorem{theorem}{Theorem}
\newtheorem*{theorem*}{Theorem}

\newtheorem{remark}{Remark}

\numberwithin{equation}{section}

\begin{document}

\preprint{}

\title{Bach equation and the matching of spacetimes in conformal cyclic cosmology models}

\author{Jaros\l aw Kopi\'nski}
 \email{jkopinski@cft.edu.pl}
\affiliation{Center for Theoretical Physics, Polish Academy of Sciences, \\ Aleja Lotnik\'ow 32/46, 02-668 Warsaw, Poland}
\author{Juan A. Valiente Kroon}
 \email{j.a.valiente-kroon@qmul.ac.uk}
\affiliation{
School of Mathematical Sciences, Queen Mary University of London, \\ Mile End Road, London E1 4NS, United Kingdom
}%

\begin{abstract}
We consider the problem of matching two spacetimes, the previous and present aeons, in the conformal cyclic cosmology model. The common boundary between them inherits two sets of constraints -- one for each solution of the Einstein field equations extended to the conformal boundaries. The previous aeon is assumed to be an asymptotically de Sitter spacetime, so the standard conformal formulation of the Einstein field equations suffices to derive the constraints on the future null infinity. For the future aeon, which is supposed to evolve from an initial singularity, they are obtained with the use of the Bach-type equation. This equation is regular at the past conformal infinity for conformally flat and conformally Einstein spacetimes, so we will mostly focus on them here. An example of the electrovacuum spacetime which does not fall into this class and has a regular conformal Bach tensor will be discussed in the Appendix.
\end{abstract}

\pacs{04.20.Ex,04.70.Bw,04.20.Jb}

\maketitle

\noindent

\section{Introduction}

The conformal cyclic cosmology (CCC) is an alternative cosmological model, introduced by Penrose \cite{penrose_ccc}. It is based on the common agreement that our Universe evolves from an initial singularity, and that this evolution is influenced by the presence of a positive cosmological constant. Hence, its end state is assumed to approach an asymptotically de Sitter spacetime. This solution of the Einstein field equation has the remarkable property that its future conformal boundary is a spacelike hypersurface, which allows one to perform matching with the spacelike past conformal boundary of the other spacetime. Such gluing procedure can be viewed as a transition between two distinct solutions of the Einstein field equations. This is the bedrock of the CCC model: postulating that the history of our Universe consists of such building blocks, aeons, which are joined together by their spacelike conformal boundaries to form an infinite cycle. Because the aeons evolve from the big bang singularity into asymptotically de Sitter spacetimes, the CCC provides a natural explanation for the Weyl curvature hypothesis \cite{penrose_wch}. One of the other physical implications of this model is the postulated existence of the ringlike structures in the cosmic microwave background \cite{cmb1, cmb2, lopez}.

There have been many approaches to the construction of a viable model of the transition between two aeons in the CCC---see e.g. \cite{newman, tod, nurek_meissner, nurowski}. However, all of them assume that the conformal metric is given either in an exact form or in terms of a power series. The present article serves as a step toward the general case. To carry out our analysis, we make use of the conformal Einstein field equations and the Bach-type equation to study the constraints on the common boundary between aeons. The assumption that the conformal version of the latter is regular on the conformal boundary is a very restrictive condition. As will be seen, a natural way to satisfy it is to consider conformally Einstein spacetimes. This will be the main focus of this work. However, a class of electrovacuum spacetimes that are not conformally flat and have a regular Bach equation will also be discussed in the Appendix as an example of a more general scenario.

\medskip
The structure of this article is as follows. In the next section, we will discuss the details of the transition between the two aeons of the CCC model. Section \ref{cefesec} provides a discussion of the constraints induced on the conformal boundary of the previous aeon by Friedrich's conformal Einstein field equations \cite{friedrich1, friedrich2}. After that, in Sec. \ref{bacheq}, the Anderson-Fefferman-Graham equation \cite{anderson, anderson_chrusciel} (which in the current case reduces to the equation with Bach tensor) and a nonlinear wave equation for the conformal factor will be used to describe the constraints on the past null infinity of the present aeon. The analysis of the big bang singularities that uses the Bach equation is viable only for a certain class of spacetimes --- most notably the conformally flat ones. However, based on an example of the electrovacuum solution of the Einstein field equations discussed in the Appendix, we will make the assumption that the Bach equation is regular everywhere to analyse the present aeon. Lastly, in Section \ref{secconc} the two sets of constraints obtained in Secs. \ref{cefesec} and \ref{bacheq} will be evaluated on the common boundary between the aeons to obtain certain simplifications.

\medskip
This article relies on the conformal formulation of the Einstein field equation, where asymptotically de Sitter spacetime $(\widehat{\mathcal{M}}, \widehat{g}_{ab})$ is assumed to have a conformal extension $(\mathcal{M}, g_{ab})$ in the form of compact manifold $\mathcal{M}$ with the boundary $\partial \mathcal{M}$ and the metric $g_{ab} = \Omega^2 \widehat{g}_{ab}$. The function $\Omega$ is positive in the interior of $\mathcal{M}$ and vanishes on $\partial \mathcal{M}$. In the analysis of the conformal field equations it is convenient to introduce the rescaled Weyl and Cotton tensors of $\widehat{g}_{ab}$ in the following way,
\begin{equation}
d^a{}_{bcd} := \frac{1}{\Omega} C^a{}_{bcd}, \quad Q_{abc} := \frac{1}{\Omega} \widehat{A}_{abc}.
\end{equation}
The spacetime with the initial (big bang) singularity will be assumed to satisfy the following regular Bach equation,
\begin{equation}
\overline{B}_{ab} = S_{ab}, 
\end{equation}
where $\overline{B}_{ab}$ is the Bach tensor of the conformal metric and $S_{ab}$ is the source term. With this preparation, we can formulate the main theorem of this article.
\begin{theorem*}
Let $(\widehat{\mathcal{M}}, \widehat{g}_{ab})$ be an asymptotically de Sitter spacetime with the cosmological constant $\widehat{\Lambda}$ and the future conformal boundary $\mathcal{I}^+$ where the Cotton tensor vanishes. Assume that $(\widecheck{\mathcal{M}}, \widecheck{g}_{ab})$ is a spacetime with the big bang singularity located on the past conformal boundary $\mathcal{I}^-$, which is a solution of the regular conformal Bach equation. Then, if $(\widehat{\mathcal{M}}, \widehat{g}_{ab})$ and $(\widecheck{\mathcal{M}}, \widecheck{g}_{ab})$ are subsequent aeons in the conformal cyclic cosmology scenario the constraints 
\begin{equation}
    \begin{split}
    & D^b e_{ab} \doteq - Q^{\top}_{ \perp \perp a} \doteq \sqrt{\frac{3}{\widehat{\Lambda}}} S^{\top}_{a \overline{\perp} } , \quad  Q^{\top}_{\perp bc } \doteq 0, \\
    & b_{bca} \doteq  \sqrt{\frac{3}{\widehat{\Lambda}}} A^{(3)}_{abc}, \quad S_{\overline{\perp} \overline{\perp}}  \doteq 0.
\end{split}
\end{equation}
have to be satisfied on the common conformal boundary between them. The quantities $e_{ab}$ and $b_{abc}$ correspond to the electric and magnetic part of the rescaled Weyl tensor $d_{abcd}$ and $A^{(3)}_{abc}$ is the Cotton tensor of the conformal boundary.
\end{theorem*}

As an application of the main theorem we derive constraints relating the initial values of the matter fields in the present aeon with the same fields from the previous aeon if both spacetimes are assumed to be electrovacuum. This provides a partial answer to one of the questions posed by Tod in \cite{todqccc}. 

\subsection*{Notation and conventions}
We will work with the 4-dimensional spacetimes and use abstract index notation throughout the paper. The signature of the spacetime metrics will be $(-,+,+,+)$. The convention for the Riemann tensor $R_{abcd}$ is as follows
\begin{equation}
\nabla_a \nabla_b v_c - \nabla_b \nabla_a v_c = R_{abcd}v^d.
\end{equation}

It can be decomposed into the Weyl tensor $C_{abcd}$ and the Schouten tensor in the following way:
\begin{equation}
R_{abcd} = C_{abcd} + 2 \left(g_{c[a}P_{b]d} + g_{d[b}P_{a]c}\right),
\end{equation}
where
\begin{equation}
P_{ab} := \frac{1}{2}R_{ab} - \frac{R}{12} g_{ab}, \quad J:= P_a{}^a = \frac{R}{6},
\end{equation}
is the trace-corrected Ricci tensor ---also called the Schouten tensor. The (anti)symmetrization brackets are defined as
\begin{equation}
T_{(ab)} = \frac{1}{2} \left(T_{ab} + T_{ba} \right), \quad T_{[ab]} = \frac{1}{2} \left(T_{ab} - T_{ba} \right).
\end{equation}
with the obvious generalizations to more indices. 

Let $\mathcal{S}$ be a codimension-one spacelike hypersurface with the induced metric $h_{ab}$ and the unit normal vector $n^a$. The projection to hypersurface tensors will be denoted by a superscript $\top$, whereas contraction with $n^a$ by $\perp$, e.g.
\begin{equation}
T^\top_{a \perp} = h_a{}^b T_{b c}n^c.
\end{equation}

The Weyl tensor can be decomposed into its electric and magnetic part with respect to an unit vector $u^a$ as follows
\begin{align} 
E_{ab} := u^c u^d C_{acbd}, \quad H_{ab} := \frac{1}{2} u^c u^d \eta_{ackl}C^{kl}{}_{bd},
\end{align}
where $\eta_{abcd}$ is the covariant Levi-Civita tensor. Lastly, the Cotton and Bach tensors are defined as 
\begin{equation}
\begin{aligned}
A_{abc} & := 2 \nabla_{[b}P_{c]a}, \\
B_{ab} & := - \nabla^c A_{abc} + P^{dc} C_{dacb}.
\end{aligned}
\end{equation}
Moreover, the Bianchi identity implies
\begin{equation}
A_{abc} = \nabla^d C_{dabc}.
\end{equation}

\section{Conformal Cyclic Cosmologies}
In the following, let $( \widehat{\mathcal{M}}, \widehat{g}_{ab} )$ and $( \widecheck{\mathcal{M}}, \widecheck{g}_{ab} )$ denote two solutions of the Einstein field equations that will be called, respectively, the previous and present aeons. One has that 
\begin{equation}
\begin{aligned}
& \widehat{R}_{ab} - \frac{1}{2} \widehat{R} \widehat{g}_{ab} + \widehat{\Lambda} \widehat{g}_{ab} = \widehat{T}_{ab}, \\
& \widecheck{R}_{ab} - \frac{1}{2} \widecheck{R} \widecheck{g}_{ab} + \widecheck{\Lambda} \widecheck{g}_{ab} = \widecheck{T}_{ab},
\end{aligned}
\end{equation}
where $ \widehat{\Lambda}$ (respectively  $\widecheck{\Lambda}$) are the positive cosmological constants and $\widehat{T}_{ab}$ (respectively $\widecheck{T}_{ab}$) are the energy-momentum tensors that describe the matter content of the corresponding aeon. We will assume that $( \widehat{\mathcal{M}}, \widehat{g}_{ab} )$ is an asymptotically de Sitter spacetime and that $( \widecheck{\mathcal{M}}, \widecheck{g}_{ab} )$ evolves from an initial singularity. Those statements can be made precise with the assumption that both spacetimes admit a conformal compactification. Hence, in the sequel we will use the notion of conformal extensions of both aeons, $(\mathcal{M}, g_{ab})$ and $(\overline{\mathcal{M}}, \overline{g}_{ab})$, such that:

\begin{itemize}
\item[(i)] $\mathcal{M}$ (respectively $\overline{\mathcal{M}}$) are compact manifolds with the boundaries $\partial \mathcal{M}$ (respectively $\partial \overline{\mathcal{M}}$) such that
\begin{equation}
\mathcal{M} = \widehat{\mathcal{M}} \cup \partial \mathcal{M}, \quad \overline{\mathcal{M}} = \widecheck{\mathcal{M}} \cup \partial \overline{\mathcal{M}}.
\end{equation}
\item[(ii)] The metrics $g_{ab}$  (respectively $\overline{g}_{ab}$) are regular everywhere and there exist positive functions $\Omega$ and $\omega$ such that
\begin{equation} 
 g_{ab} = \Omega^2 \widehat{g}_{ab}, \qquad \overline{g}_{ab} = \frac{1}{\omega^2} \widecheck{g}_{ab}.
\end{equation}
That is, the zero locus $\mathcal{Z} (\Omega)$ corresponds to the future null infinity (future conformal boundary) $\mathcal{I}^+$ of $( \widehat{\mathcal{M}}, \widehat{g}_{ab} )$ and $\mathcal{Z} (\omega)$ to the past conformal boundary $\mathcal{I}^-$ of  $( \widecheck{\mathcal{M}}, \widecheck{g}_{ab} )$.

\end{itemize}

It is important to note here that we are not assuming that both spacetimes have the same conformal extension. If that was the case, then the reciprocal hypothesis could be used to determine the metric of the present aeon out of the metric of the past one, given that the conformal factor could be prescribed uniquely, or vice versa \cite{penrose_ccc, tod}. Instead, we will work with the two solutions of the Einstein field equations which independently satisfy the required conformal properties and study the matching conditions along their conformal boundary afterwards.

\section{The asymptotically de Sitter aeon: Friedrich's conformal field equations and constraints} \label{cefesec}
The objective of this section is to consider asymptotically de Sitter-like spacetimes $( \widehat{\mathcal{M}}, \widehat{g}_{ab} )$ which admit conformal extensions $(\mathcal{M}, g_{ab})$ with
\begin{equation}
g_{ab} = \Omega^2 \widehat{g}_{ab}
\end{equation}
and derive the constraints on the future conformal boundary $\mathcal{I}^+$ ($\mathcal{Z}(\Omega)$). This is a standard setting for Friedrich's conformal Einstein field equations \cite{friedrich1, friedrich2} and we will briefly describe this approach with nonvanishing energy-momentum tensor ---see \cite{javk} and the references therein for the discussion of trace-free matter models.

\subsection{Conformal Einstein field equations} 
By the conformal Einstein field equations it is understood a conformal representation of the Einstein field equations ---that is, they provide a set of equations which are formally regular at the conformal boundary (where the conformal factor $\Omega$ vanishes) of an asymptotically de Sitter spacetime, which imply, away from the conformal boundary, a solution to the Einstein field equations. There exist in the literature several candidates for suitable conformal field equations --- see e.g. \cite{javk}. In this article we focus on Friedrich's conformal Einstein field equations \cite{Fri83}. These equations, which in the following we just simply call \emph{the conformal Einstein equations}, have been instrumental in the study of the nonlinear stability of de Sitter-like and Minkowski-like spacetimes \cite{Fri86}. Although the conformal Einstein equations have mostly been considered in the vacuum setting, the formalism can be extended to the nonvacuum case. Indeed, these equations have been used to study the stability of a number of cosmological models (e.g. scalar field, dust, radiation) \cite{Fri17,LueVal13,CarHurVal18}.  As will be seen in the following, these equations allow the formulation of a regular asymptotic initial value problem at the conformal boundary.

\medskip
As mentioned above, the strategy behind the conformal Einstein field equations is to find a system that is regular at the conformal boundary $\mathcal{I}^+$ and is equivalent to the usual Einstein field equation under suitable conditions. Let
\begin{equation}
s := \frac{1}{24} R \Omega + \frac{1}{4} \Box \mspace{4mu} \Omega
\end{equation}
where the curvature quantities and derivative operator is associated with the conformal metric $g_{ab}$. It is also useful to define the rescaled version of the tensors corresponding to the de Sitter-like (physical) metric $\widehat{g}_{ab}$,
\begin{equation}
d^a{}_{bcd} := \frac{1}{\Omega} C^a{}_{bcd}, \qquad Q_{abc} := \frac{1}{\Omega} \widehat{A}_{abc},
\end{equation}
The conformal Einstein field equations are given by the following system:
\begin{subequations} \label{cefe}
\begin{align}
& \nabla_a \nabla_b \Omega  = - \Omega P_{ab} + s g_{ab} + \frac{\Omega}{2} \mathring{T}_{ab}, \label{cefe1} \\
 & \nabla_a s  = -P_{a}{}^b \nabla_b \Omega + \frac{1}{6} \nabla^c \left( \mathring{T}_{ac} \Omega \right) , \label{cefe2} \\
 & \nabla_d d^d{}_{abc}  = Q_{abc}, \label{cefe3} \\
 & 2 \nabla_{[b} P_{c]a}  = d^d{}_{abc} \nabla_d \Omega + \Omega Q_{abc}, \label{cefe4} \\
 & \widehat{\Lambda}  = 6 \Omega s - 3 \nabla_c \Omega \nabla^c \Omega + \frac{\widehat{T}}{4}. \label{cefe5}
\end{align}
\end{subequations}
where $ \mathring{T}_{ab}$ denotes the trace-free part of the physical energy-momentum tensor 
\begin{equation}
 \mathring{T}_{ab} = \widehat{T}_{ab} -\frac{\widehat{T}}{4} \widehat{g}_{ab}, 
\end{equation}
further 
complemented by the conservation equation for $ \widehat{T}_{ab}$ and its conformal transformation rule 
\begin{equation}\label{phystrule}
T_{ab} = \frac{1}{\Omega^q} \widehat{T}_{ab}
\end{equation}

where $T_{ab}$ is the regular unphysical energy-momentum tensor. Because the spacetime is asymptotically de Sitter we will restrict ourselves to $q \geq 0$.

 A solution to the system \eqref{cefe} has the form of a collection of fields
\begin{equation}
\left(\Omega, g_{ab}, s, P_{ab}, d^a{}_{bcd}, T_{ab} \right)
\end{equation}
over $\mathcal{M}$.

\begin{remark} {\em As $\nabla_a \Omega$ is orthogonal to level set of $\Omega$, Eq. (\ref{cefe5}) is equivalent to the statement that $\mathcal{Z} \left( \Omega \right)$ is a spacelike hypersurface ---as long as $\widehat{T}=0$ there.}
\end{remark} 

\begin{remark} \label{remcons}
{\em If the trace $\widehat{T}$ vanishes, then for $q=2$ the conservation equation $\widehat{\nabla}_b \widehat{T}_a{}^b=0$ is conformally invariant and regular everywhere.}
\end{remark} 

\subsection{Constraints}
Let $\mathcal{S}$ be a spacelike hypersurface of $\mathcal{M}$ with unit normal vector $n^a$, $n^a n_a = -1$. The unphysical metric $g_{ab}$ induces a metric $h_{ab}$ on $\mathcal{S}$, given by
\begin{equation}
    h_{ab} := g_{ab} + n_a n_b.
\end{equation}
The extrinsic curvature of $\mathcal{S}$ will be denoted by $K_{ab}$ and is defined as
\begin{equation}
    K_{ab} := \nabla_{a} n_b + n_a a_b,
\end{equation}
where $a_b = \nabla_n n_b$ is the acceleration of $n^a$. Let $\sigma := \nabla_{\perp} \Omega$, so that one has the decomposition
\begin{equation}
    \nabla_a \Omega = D_a \Omega - n_a \sigma,
\end{equation}
where $D_a$ is the Levi-Civita connection of $h_{ab}$.

\medskip
Before discussing the constraints induced by the conformal field equations on $\mathcal{S}$, it is useful to consider the decompositions of the relevant tensors with respect to $n^a$. We have
\begin{equation}
        \mathring{T}_{ab} = \mathring{T}_{ab}^{\top} - 2  n_{(a} \mathring{T}_{ b) \perp}^{\top}  +  n_a n_b \mathring{T}_{\perp \perp},
\end{equation}
Similarly,
\begin{equation}
    P_{ab} = P_{ab}^{\top} - 2  n_{(a} P_{ b) \perp}^{\top}  +  n_a n_b P_{\perp \perp}.
\end{equation}
The decomposition of the rescaled Weyl tensor $d_{abcd}$ is given by
\begin{equation}
\begin{split}
    d_{abcd}&  =   d^{\top}_{abcd} -  n_d b_{abc} + n_c b_{abd} - n_b b_{cda} \\
    & + n_a b_{cdb} + n_a n_c e_{bd} + n_b n_d e_{ac} \\
    & - n_b n_c e_{ad} - n_a n_d e_{bc},
\end{split}
\end{equation}
where $e_{ab} :=  d^{\top}_{a \perp b \perp}$ is its electric part and $b_{abc} := d^{\top}_{abc\perp}$ defines its magnetic part. Because of the symmetries of $d_{abcd}$ its fully projected part can be further decomposed as
\begin{equation}
    d^{\top}_{abcd} = e_{ac} h_{bd} + e_{bd} h_{ac}  - e_{ad} h_{bc} - e_{bc} h_{ad}.
\end{equation}

\medskip
The constraints induced on $\mathcal{S}$ by \eqref{cefe} can be obtained by considering intrinsic and normal-intrinsic components of this equations. They read
\begin{subequations} \label{cefec}
\begin{align}
 D_a D_b \Omega & =  \sigma K_{ab} - \Omega P_{ab}^{\top} + s h_{ab} + \frac{\Omega}{2} \mathring{T}_{ab}^{\top}, \\
 D_{a} \sigma & = K_{a}{}^b D_b \Omega - \Omega P_{ a \perp}^{\top} + \frac{\Omega}{2}\mathring{T}_{ a \perp}^{\top},  \\
\begin{split}
  D_a s  & = -  P_{ab}^{\top} D^b \Omega + P_{ a \perp}^{\top}\sigma  +\frac{1}{6} \mathring{T}_{ab}^{\top} D^b \Omega \\ & - \frac{1}{6} \mathring{T}_{ a \perp}^{\top} \sigma  +  \frac{\Omega}{6} \bigg( D^b \mathring{T}_{ab}^{\top} +\mathring{T}_{ab}^{\top} a^b  \\
   & - K_{a}{}^b \mathring{T}_{ b \perp}^{\top} - K \mathring{T}_{ a \perp}^{\top}  - h_a{}^b \nabla_{\perp} \mathring{T}_{ b \perp}^{\top}  \\
    & +  \mathring{T}_{ \perp \perp} a_a  \bigg),
\end{split} \\
   D_b e^b{}_a &  = - b_{bac} K^{bc} -  Q^{\top}_{\perp \perp a}, \\
 D^a b_{bca} & = 2  e_{a[b}K^a{}_{c]}   +  Q^{\top}_{\perp bc}, \\
 \begin{split}
  2 D_{[b} P^{\top}_{c]a} &  = - 2  P^{\top}_{ \perp [b } K_{c]a}  + \Omega  Q^{\top}_{abc}  + \sigma b_{bca} \\
  & - 2  h_{a[b} e_{c] d}D^d \Omega - 2 e_{a[b} D_{c]}\Omega,
 \end{split} \\
  2 D_{[a} P^{\top}_{b] \perp} & = 2 K_{[a}{}^c P^{\top}_{b]c} + b_{abc}D^c \Omega + \Omega   Q^{\top}_{\perp ab}, \\
  \widehat{\Lambda} & = 6 \Omega s  - 3 D_a \Omega D^a \Omega + 3  \sigma^2 + \frac{\widehat{T}}{4},
\end{align}
\end{subequations}
where $K:=K_a{}^a$. Any other projections are either trivial or can be expressed by the linear combination of (\ref{cefec}). This system of equations is supplemented by
    \begin{equation}
    \begin{split}
        P^{(3)}_{ab} & = \Omega e_{ab} + P^{\top}_{ab} - K \left( K_{ab} - \frac{1}{4} h_{ab} K \right) \\
        & + K_{ac} K_b{}^c - \frac{1}{4} h_{ab} K_{cd} K^{cd},
    \end{split}
    \end{equation}
where $P^{(3)}_{ab}$ is the Schouten tensor of $h_{ab}$, and
\begin{equation}
    2 D_{[a}K_{b]c} = \Omega b_{abc} + 2 h_{c[a}P^{\top}_{b] \perp}.
\end{equation}
These two equations are a consequence of the Gauss-Codazzi and Codazzi–Mainardi relations.

\subsection{Constraints on $\mathcal{Z} \left( \Omega \right)$} 
Now, suppose that $\mathcal{S}$ is the future conformal boundary of the asymptotically de Sitter spacetime ---i.e. $\mathcal{S} = \mathcal{Z} \left( \Omega \right)$. The constraint equations (\ref{cefec}) reduce greatly in that case and read
\begin{subequations}
\begin{align}
	D_a \sigma & \doteq 0,\quad \widehat{\Lambda}  \doteq  3  \sigma^2 , \label{csig1}  \\
    s h_{ab} & \doteq - \sigma K_{ab}, \label{umbilicp} \\
    \quad D_a s & \doteq   \sigma \left(  P^{\top}_{a \perp} - \frac{1}{6} \mathring{T}^{\top}_{a \perp}\right), \label{das} \\
    \begin{split}
    P^{(3)}_{ab} & \doteq  P^{\top}_{ab} - K \left( K_{ab} - \frac{1}{4} h_{ab} K \right) \\
    & + K_{ac} K_b{}^c - \frac{1}{4} h_{ab} K_{cd} K^{cd},
    \end{split}  \label{pabcondef} \\
    D_{[a}K_{b]c} & \doteq   h_{c[a}P^{\top}_{b] \perp}, \label{j0eq} \\
    2 D_{[a} P^{\top}_{b] \perp }  & \doteq 2 K_{[a}{}^c P^{\top}_{b]c}, \label{pbeq} \\
    2 D_{[b} P^{\top}_{c]a} & \doteq   \sigma b_{bca} - 2  P^{\top}_{\perp [b} K_{c]a}, \label{bdef} \\
    D^b e_{ab} & \doteq -  Q^{\top}_{\perp \perp a } - K^{bc} b_{bac},  \label{con1} \\
    D^a b_{bca} & \doteq  Q^{\top}_{\perp bc} - 2  K_{[b}{}^a e_{c]a}, \label{con2}  
\end{align}
\end{subequations}
where $\doteq$ denotes the equality on the zero set of $\Omega$. Equations \eqref{csig1} yield $\sigma \doteq \sqrt{\widehat{\Lambda}/3}$. Because this function (the normal derivative of $\Omega$) is constant on $\mathcal{Z} \left( \Omega \right)$ it is convenient to define
\begin{equation}
\kappa := \frac{s}{\sigma}.
\end{equation}
Then, from (\ref{umbilicp}) and (\ref{das})
\begin{equation}
K_{ab} \doteq - \kappa h_{ab}, \quad  P^{\top}_{a \perp} \doteq  D_a \kappa + \frac{1}{6} \mathring{T}^{\top}_{a \perp},
\end{equation}
e.g. $\mathcal{Z} \left( \Omega \right)$ is an umbilic hypersurface. The Gauss-Codazzi constraint \eqref{pabcondef} may serve as a definition of the projected part of the unphysical Schouten tensor. More precisely, one has
\begin{equation}
    P^{\top}_{ab} \doteq P^{(3)}_{ab} + \frac{1}{2} \kappa^2 h_{ab}.
\end{equation}
It can be readily verified that the Codazzi-Mainardi relation constraint \eqref{j0eq} yields $\mathring{T}^{\top}_{a \perp} \doteq 0$ in the current case, which leads to \eqref{pbeq} being identically satisfied.

\medskip
The constraint \eqref{bdef} may now be viewed as a definition of the magnetic part of the rescaled Weyl tensor, i.e.
\begin{equation} \label{bsch3}
      b_{bca} \doteq  \frac{1}{\sigma} 2 D_{[b}P^{(3)}_{c]a} = \frac{1}{\sigma} A^{(3)}_{abc},
\end{equation}
 where $A^{(3)}_{abc}$ is the unphysical 3-dimensional Cotton tensor. The last two constraints, (\ref{con1}) and (\ref{con2}), reduce to
\begin{equation} \label{ebcon}
    D^b e_{ab} \doteq - Q^{\top}_{ \perp \perp a}, \quad  Q^{\top}_{\perp ab } \doteq 0,
\end{equation}
where the divergence-free nature of $A^{(3)}_{abc}$ has been used.

\medskip
This section can be summarized as follows:
\begin{theorem}
Let $(\widehat{\mathcal{M}}, \widehat{g}_{ab})$ be an asymptotically de Sitter spacetime with conformal extension $(\mathcal{M},g_{ab})$ where $g_{ab}=\Omega^2 \widehat{g}_{ab}$. Then, the constraints induced on the future conformal boundary $\mathcal{I}^+$ (zero locus of $\Omega$) are as follows:
\begin{equation} \label{constrpreviou}
\begin{split}
& \nabla_\perp \Omega \doteq \sqrt{\frac{\widehat{\Lambda}}{3}}, \quad  K_{ab} \doteq -  \sqrt{\frac{3}{\widehat{\Lambda}}} s h_{ab} \\
&P^{\top}_{a \perp} \doteq  \sqrt{\frac{3}{\widehat{\Lambda}}} D_{a}s, \quad   P^{\top}_{ab } \doteq P^{(3)}_{ab} + \frac{3}{2 \widehat{\Lambda} }  s^2 h_{ab}, \\
&  b_{abc} \doteq \sqrt{\frac{3}{\widehat{\Lambda}}} A^{(3)}_{cab}, \quad  D^b e_{ab} \doteq - Q^{\top}_{ \perp \perp a}, \\
&   Q^{\top}_{\perp ab } \doteq 0,
\end{split}
\end{equation}
where the unhatted quantities correspond to the conformal metric $g_{ab}$ and the superscript $(3)$ indicates quantities intrinsic to $\mathcal{I}^+$. Moreover, $s := \frac{1}{24} R \Omega + \frac{1}{4} \Box \mspace{4mu}  \Omega$, $e_{ab}$ and $b_{abc}$ correspond to the electric and magnetic part of the rescaled Weyl tensor $\Omega^{-1} C_{abcd}$ and $Q_{abc}:= \Omega^{-1} \widehat{A}_{abc}$ is the rescaled Cotton tensor. 
\end{theorem}

\begin{remark}
{\em Following the postulates of the CCC, the conformal boundary is the natural hypersurface to perform matching between a past aeon and a future one.}
\end{remark}

\section{The Big Bang singularity: the present aeon} \label{bacheq}
Having described the constraints on the future conformal boundary of the asymptotically de Sitter (previous) aeon we will move to the formulation of the conformally regular version of the Einstein field equations for the present aeon. As will be seen, the approach described in the previous section will not work due to the requirement of an initial singularity.

\medskip
Recall that $(\widecheck{\mathcal{M}}, \widecheck{g}_{ab})$ is a solution of the Einstein field equations with the energy-momentum tensor $\widecheck{T}_{ab}$ and the positive cosmological constant $\widecheck{\Lambda}$. Moreover, we assume the existence of a conformal extension of this spacetime $(\overline{\mathcal{M}}, \overline{g}_{ab})$ such that
\begin{equation}
 \overline{g}_{ab} = \frac{1}{\omega^2} \widecheck{g}_{ab} 
\end{equation}
and $\mathcal{Z} \left( \omega \right)$ corresponds to the initial (big bang) singularity. Since the physical energy-momentum tensor is singular at $\mathcal{Z} \left( \omega \right)$ let
\begin{equation} \label{unphystcheck}
    \widecheck{T}_{ab} =   \omega ^{-v} \overline{T}_{ab} \implies \widecheck{T} = \omega^{-2- v} \overline{T},
\end{equation}
where $v \geq 0$ and $\overline{T}_{ab}$ (the unphysical energy-momentum tensor) is regular everywhere. The Einstein field equations can be written as
\begin{equation} \label{efepresent}
    \widecheck{P}_{ab} = \frac{1}{2} \widecheck{T}_{ab} + \frac{1}{6} \widecheck{g}_{ab} \left( \widecheck{\Lambda} - \widecheck{T} \right).
\end{equation}
In terms of conformal quantities, these last expressions implies 
\begin{equation} \label{iefeconf}
\begin{split}
 &   \omega^2 \overline{P}_{ab} + 2 \overline{\nabla}_a \omega \overline{\nabla}_b \omega - \frac{1}{2} \overline{g}_{ab} \overline{\nabla}_c \omega \overline{\nabla}^c \omega - \omega \overline{\nabla}_a \overline{\nabla}_b \omega \\
   &  = \frac{1}{2} \omega^{2- v} \left( \overline{T}_{ab} - \frac{1}{3} \overline{g}_{ab} \overline{T} \right) + \frac{1}{6} \overline{g}_{ab} \omega^4 \widecheck{\Lambda}. 
\end{split}
\end{equation}
\begin{remark}
{\em Equation \eqref{iefeconf} can be used to derive regularity conditions on the unphysical energy-momentum tensor $\overline{T}_{ab}$ at $\mathcal{Z} \left( \omega \right)$. After taking its trace one obtains
\begin{equation}
\omega^{2- v} \overline{T} \doteq 0.
 \end{equation}
Similarly, if $\overline{n}_a := \overline{\nabla}_a \omega$ is a normal vector to the level set of $\omega$, then
\begin{equation}
  2 \overline{n}_a  \overline{n}_b - \frac{1}{2} \overline{g}_{ab} \overline{n}_c  \overline{n}^c \doteq \frac{1}{2} \omega^{2- v} \left( \overline{T}_{ab} - \frac{1}{3} \overline{g}_{ab} \overline{T} \right)  \bigg|_{\omega = 0 }.
\end{equation}
If $\mathcal{Z} \left( \omega \right)$ is a spacelike hypersurface, then after contracting the equality above with the metric $\overline{h}_{ab}$ (intrinsic to  $\mathcal{Z} \left( \omega \right)$) one gets
\begin{equation}
 \omega^{2- v}  \overline{T}_{\overline{\perp} \overline{\perp} }   \bigg|_{\omega = 0 } \geq 0.
\end{equation}
}
\end{remark}

Because of the trace-free character of the $\omega^{-2}$ terms in Eq. \eqref{efepresent} written in terms of conformal quantities, a similar regularization procedure as Friedrich's conformal Einstein field equations cannot be employed here to obtain a system that is regular on $\mathcal{Z} \left( \omega \right)$. However, in the sequel we will argue that if the physical energy-momentum tensor of the matter is restricted in a certain way, then an approach that uses the Bach tensor can be employed to produce a suitable set of regular conformal equations. 

\subsection{The Bach equation}
 The Bach tensor $\widecheck{B}_{ab}$ can be related to the energy-momentum tensor $\widecheck{T}_{ab}$ with the use of the Einstein field equations. We have
\begin{equation} \label{bacht}
\begin{split}
    \widecheck{B}_{ab} & = \frac{1}{2} \widecheck{\Box} \mspace{4mu}  \widecheck{T}_{ab} - \frac{1}{6} \widecheck{g}_{ab} \widecheck{\Box} \mspace{4mu}  \widecheck{T} + \frac{1}{6} \widecheck{\nabla}_a \widecheck{\nabla}_b \widecheck{T} - \widecheck{T}_{a}{}^c \widecheck{T}_{bc}  \\
    & + \frac{1}{6} \widecheck{g}_{ab} \widecheck{T} \left( \widecheck{\Lambda} - \widecheck{T} \right) +  \widecheck{C}_{acbd} \widecheck{T}^{cd} + \frac{1}{4} \widecheck{g}_{ab} \widecheck{T}_{cd} \widecheck{T}^{cd} \\
    & - \frac{2}{3} \widecheck{T}_{ab} \left( \widecheck{\Lambda} - \widecheck{T} \right).
\end{split}
\end{equation}
It can be verified that for $ \widecheck{g}_{ab} \mapsto \overline{g}_{ab} = \omega^{-2} \widecheck{g}_{ab}$ the Bach tensor transforms as
\begin{equation}
    \widecheck{B}_{ab} \mapsto \overline{B}_{ab} = \omega^2 \widecheck{B}_{ab},
\end{equation}
so the initial step in obtaining the conformal field equations which describe an initial singularity spacetime is to verify whether the right-hand side of (\ref{bacht}) multiplied by $\omega^2$ and written in terms of the unphysical energy-momentum tensor is regular on $\mathcal{Z} \left( \omega \right)$. Unfortunately, a simple calculation reveals that this is not the case for generic $\overline{T}_{ab}$. Nevertheless, driven by an example concerning certain class of electrovacuum spacetimes (as discussed in the Appendix) where the the source term in the Bach equation \emph{is} regular everywhere, we will proceed with the analysis of the constraints assuming that the conformal equation has the form 
\begin{equation} \label{bachreg}
\overline{B}_{ab} = S_{ab},
\end{equation}
where $S_{ab}$ is a (regular) source term. It should be stressed, however, that although Eq. \eqref{bachreg} bears close resemblance to the equation appearing in the conformally invariant theories of gravity (see \cite{mannheim} for a recent developments), here we take the point of view that it is a consequence of the Einstein field equations.

\begin{remark}
{\em The source term $S_{ab}$ vanishes for conformally flat and conformally Einstein spacetimes.}
\end{remark}

The Bach tensor is equivalent, in dimension 4, to the Fefferman-Graham obstruction tensor --- see \cite{feff_gr}. Hence, an approach based on equation $B_{ab} = 0$ has been used to generalize Friedrich's approach to higher-dimensional asymptotically de Sitter spacetimes \cite{anderson, anderson_chrusciel} ---see also \cite{kaminski}. In our analysis the conformal extension of the present aeon is a solution of the Bach equation with a source term, so a similar approach can be used to show the well-posedness of this system. The relation between its solution and the physical big bang singularity spacetime can be achieved with the use of the nonlinear wave-type equation for the conformal factor $\omega$.

\subsection{The regularized Bach equation constraints on the conformal boundary $\mathcal{I}^-$}
Following the discussion in the previous subsection we will assume that the conformal version of the Einstein field equations of $(\widecheck{\mathcal{M}}, \widecheck{g}_{ab})$ are given by
\begin{equation} \label{confpresent}
\begin{split}
    \overline{B}_{ab} & = S_{ab}, \\
     \widecheck{\Box} \mspace{4mu}  \omega & = \overline{J} \omega - \frac23 \omega^3 \widecheck{\Lambda} + \frac16 \overline{T} \omega^{1-v},
\end{split}
\end{equation}
where $S_{ab}$ is a regular source term which depends on the matter content of the spacetime and the wave-type equation for the conformal factor $\omega$ arises from taking a trace of (\ref{iefeconf}).

\medskip
 The constraints on the initial hypersurface implied by the Bach equation can be obtained from the normal-normal and normal-intrinsic components of Eq. \eqref{bachreg}. More precisely, one has that
 \begin{equation}
     \overline{B}_{\overline{\perp} \overline{\perp}}= S_{\overline{\perp} \overline{\perp}}, \quad \overline{B}^{\overline{\top} }_{ a \overline{\perp} } =  S^{\overline{\top} }_{a \overline{\perp} }.
 \end{equation}
In a gauge where $\overline{\nabla}_{\overline{\perp}} \overline{n}_a =0$ on $\mathcal{Z} (\omega)$ we have (see e.g. \cite{schimming})

\begin{equation} \label{bachcons}
\begin{split}
    \overline{B}_{\overline{\perp} \overline{\perp} } & \doteq \overline{D}^a \overline{D}^b  E_{ab} - \overline{D}^c \left(C^{\overline{\top}}_{cab \overline{\perp}} \overline{K}^{ab} \right) \\ 
    & + \overline{P}^{\overline{\top}}_{ab} E^{ab} + \overline{K}^{ab} \overline{A}^{\overline{\top}}_{a \overline{\perp} b}, \\
    \overline{B}^{\overline{\top}}_{a  \overline{\perp}} & \doteq -  \overline{D}^b \overline{A}^{\overline{\top}}_{a \overline{\perp} b} - \left( \overline{D}^b E_{cb} \right) \overline{K}_a{}^c \\
     & + \overline{K}_a{}^d \overline{K}^{bc} C_{ \overline{\perp} bc d} + \overline{P}^{\overline{\top}}{}^{cd}  C^{\overline{\top}}_{d a c \overline{\perp}} + \overline{P}^{\overline{\top}}_{\overline{\perp} b} E_{a}{}^b
\end{split}
\end{equation}
where $\overline{h}_{ab} := \overline{g}_{ab} + \overline{n}_a \overline{n}_b$ and $\overline{K}_{ab}$ are the first two fundamental forms of $\mathcal{Z} \left( \omega \right)$ and $\overline{D}_a$ the Levi-Civita connection of $\overline{h}_{ab}$. Moreover, $E_{ab} = C_{a \overline{\perp}b \overline{\perp}}$ is the electric part of the Weyl tensor. To avoid introducing a new notation symbol $\doteq$ will denote equality on $\mathcal{Z}(\omega)$ here.

The boundary conditions for the nonlinear equation for the conformal factor $\omega$ consist of prescribing $\omega = 0$ on the past conformal boundary $\mathcal{I}^{-}$ and the condition that its gradient has a negative unit length there (which corresponds to $\mathcal{I}^{-}$ being a spacelike hypersurface).

\medskip
The discussion in the previous paragraphs can be summarised as follows.

\begin{theorem}
Let $(\widecheck{\mathcal{M}}, \widecheck{g}_{ab})$ be a spacetime with conformal extension $(\overline{\mathcal{M}}, \overline{g}_{ab})$ described by the conformal field equations
\begin{equation}
\begin{split}
    \overline{B}_{ab} & = S_{ab}, \\
     \overline{\Box} \mspace{4mu}  \omega & = \overline{J} \omega - \frac23 \omega^3 \widecheck{\Lambda} + \frac16 \overline{T} \omega^{1-v},
\end{split}
\end{equation}
where $\overline{g}_{ab} = \omega^{-2} \widecheck{g}_{ab}$ and $S_{ab}$ is a regular tensor which depends on the matter content on the spacetime. 
Moreover, the initial conditions for the conformal factor $\omega$ are as follows:
\begin{equation}
\omega = 0, \quad \overline{g}^{ab} \overline{\nabla}_a \omega \overline{\nabla}_b \omega = -1 \quad \mathrm{on} \quad \mathcal{I}^- \left( \widecheck{\mathcal{M}} \right),
\end{equation}
i.e. $\mathcal{I}^- \left( \widecheck{\mathcal{M}} \right) \equiv \mathcal{Z} \left( \omega \right)$ and this hypersurface is spacelike. Then, the constraints induced on the past conformal boundary $\mathcal{I}^- \left( \widecheck{\mathcal{M}} \right)$ by the first equation are
\begin{equation}
\begin{split}
 S_{\overline{\perp} \overline{\perp}} = & \overline{D}^a \overline{D}^b  E_{ab} - \overline{D}^c \left(C^{\overline{\top}}_{cab \overline{\perp}} \overline{K}^{ab} \right) \\
  & + \overline{P}^{\overline{\top}}_{ab} E^{ab} + \overline{K}^{ab} \overline{A}^{\overline{\top}}_{a \overline{\perp} b} , \\
  S^{\overline{\top}}_{a  \overline{\perp} } & =   -  \overline{D}^b \overline{A}^{\overline{\top}}_{a \overline{\perp} b} - \left( \overline{D}^b E_{cb} \right) \overline{K}_a{}^c \\
  & + \overline{K}_a{}^d \overline{K}^{bc} C_{ \overline{\perp} bc d} 
     + \overline{P}^{\overline{\top}}{}^{cd}  C^{\overline{\top}}_{d a c \overline{\perp}} + \overline{P}^{\overline{\top}}_{\overline{\perp} b} E_{a}{}^b ,
\end{split}
\end{equation}
where $E_{ab} := C_{\overline{\perp} a  \overline{\perp}  b}$ is the electric part of the Weyl tensor. 
\end{theorem}

\section{Matching conditions for the conformal boundary between aeons} \label{secconc}
In the previous sections, we discussed two sets of constraints on conformal boundaries. The first one holds on the future conformal boundary $\mathcal{I}^+$ of the previous aeon $(\widehat{\mathcal{M}}, \widehat{g}_{ab})$ (asymptotically de Sitter spacetime) and the second on the past conformal boundary $\mathcal{I}^-$ (the big bang singularity) of the present aeon $(\widecheck{\mathcal{M}}, \widecheck{g}_{ab})$. In the CCC scenario, these two hypersurfaces are identified and form a common boundary between the aeons. In the sequel, we will study the consequences of this assumption, i.e. use the constraints induced on the future conformal boundary of the previous aeon to simplify the Bach equation constraints on the past conformal boundary of the present aeon.

\medskip
The constraints coming from the conformal Einstein field equations on the future conformal boundary of the previous aeon imply that $\mathcal{I}^+$ is an umbilic hypersurface and the projections of $P_{ab}$ can be expressed in terms of $P^{(3)}_{ab}$, $h_{ab}$ and $D_a s$ as in \eqref{constrpreviou}. If we use this information in the Bach equation constraints \eqref{bachcons}, then
\begin{equation} \label{cccco}
\begin{split}
    S_{\overline{\perp} \overline{\perp}} & \doteq \overline{D}^a \overline{D}^b  E_{ab}  + \overline{P}^{(3)}_{ab} E^{ab} \\
    S^{\overline{\top}}_{a \overline{\perp} } & \doteq -  \overline{D}^b \overline{A}^{\overline{\top}}_{a \overline{\perp} b} + \sqrt{\frac{3}{\widehat{\Lambda}}} s \overline{D}^b E_{ab} + \sqrt{\frac{3}{\widehat{\Lambda}}} E_a{}^b \overline{D}_b s . 
\end{split}
\end{equation}

\medskip
where the fact that $C^{\overline{\top}}_{abc \overline{\perp}}$ vanishes on the umbilic hypersurface has been used.

A further simplification occurs after assuming that the physical Cotton tensor vanishes at $\mathcal{I}^+$, e.g. $q \geq 1$ in the conformal transformation rule (\ref{phystrule}). Then $C_{abcd} \doteq 0$ (see e.g. \cite{bonga} and Theorem 10.3 in \cite{javk}). Moreover, on the level set of $\Omega$ an equality $A_{abc} = \widehat{A}_{abc} - \sigma d_{\perp abc}$ holds, so in that case
\begin{equation}
    A^{\top}_{ab \perp} \doteq \sqrt{\frac{\widehat{\Lambda}}{3}} e_{ab}.
\end{equation}

Hence, the Bach equation constraints on the transition hypersurface between the aeons in the CCC reduce now to
\begin{equation}
\begin{split}
    S_{\overline{\perp} \overline{\perp}} & \doteq 0, \\
    S^{\overline{\top}}_{a \overline{\perp} } & \doteq  \sqrt{\frac{\widehat{\Lambda}}{3}}  \overline{D}^b e_{ab}
\end{split}
\end{equation}

The main result of this paper can be now stated in the form of the full set of the matching conditions that are imposed on the matching hypersurface between the aeons in the CCC scenario with the regular Bach equation.

\begin{theorem} \label{mth}
Let $(\widehat{\mathcal{M}}, \widehat{g}_{ab})$ be an asymptotically de Sitter spacetime with cosmological constant $\widehat{\Lambda}$ and the future conformal boundary $\mathcal{I}^+$ where the Cotton tensor vanishes. Assume that $(\widecheck{\mathcal{M}}, \widecheck{g}_{ab})$ is a spacetime with big bang singularity located on the past conformal boundary $\mathcal{I}^-$, which is a solution of the regular conformal Bach equations. Then, if $(\widehat{\mathcal{M}}, \widehat{g}_{ab})$ and $(\widecheck{\mathcal{M}}, \widecheck{g}_{ab})$ are subsequent aeons in the conformal cyclic cosmology scenario, the constraints
\begin{equation} \label{mthcons}
    \begin{split}
    & D^b e_{ab} \doteq - Q^{\top}_{ \perp \perp a} \doteq \sqrt{\frac{3}{\widehat{\Lambda}}} S^{\top}_{a \overline{\perp} } , \quad  Q^{\top}_{\perp bc } \doteq 0, \\
    & b_{bca} \doteq  \sqrt{\frac{3}{\widehat{\Lambda}}} A^{(3)}_{abc}, \quad S_{\overline{\perp} \overline{\perp}}  \doteq 0
\end{split}
\end{equation}
have to be satisfied on the common conformal boundary between $(\widehat{\mathcal{M}}, \widehat{g}_{ab})$ and $(\widecheck{\mathcal{M}}, \widecheck{g}_{ab})$. The quantities $e_{ab}$ and $b_{abc}$ correspond to the electric and magnetic part of the rescaled Weyl tensor $d_{abcd}$, $Q_{abc}$ is the rescaled Cotton tensor and $A^{(3)}_{abc}$ is the Cotton tensor of the conformal boundary. Tensor $S_{ab}$ denotes a source term in the regular Bach equation. 
\end{theorem}

\begin{remark}
{\em The source term $S_{ab}$ in the Bach equation is regular everywhere by assumption. Hence, the constraints on the common boundary between the aeons imply that the divergence of the electric part of the rescaled Weyl tensor $e_{ab}$ and the $Q^{\top}_{\perp \perp a}$ component of the rescaled Cotton tensor are regular there.}
\end{remark}

\subsection{Initial values of the (electro)magnetic fields}
One of the questions posed in \cite{todqccc} is related to the initial value of the magnetic field at the conformal boundary of the present aeon $(\widecheck{\mathcal{M}}, \widecheck{g}_{ab})$. We will show that the constraints obtained in Theorem \ref{mth} can be used to partially determine this value from the matter content of the previous aeon.

Suppose that $(\widecheck{\mathcal{M}}, \widecheck{g}_{ab})$ is the electrovacuum spacetime which satisfies the regularized Bach equations. Then,
\begin{equation}
    \overline{B}_{ab} = X_{ab} + \frac{2}{3} \widecheck{\Lambda} \omega^2 \widecheck{T}_{ab},
\end{equation}
(see the Appendix), where $X_{ab}$ is a regular tensor on the conformal extension $(\overline{\mathcal{M}}, \overline{g}_{ab})$. If we assume that $\overline{T}_{ab} = \omega^2 \widecheck{T}_{ab}$, then constraints (\ref{mthcons}) will be equivalent with
\begin{equation} \label{init1}
    \frac{2}{3} \widecheck{\Lambda} \overline{T}_{\overline{\perp} \overline{\perp} } + X_{\overline{\perp} \overline{\perp}} \doteq 0
\end{equation}
and 
\begin{equation} \label{init2}
    \frac{2}{3} \widecheck{\Lambda} \overline{T}^{\top}_{\overline{\perp} a} + X^{\top}_{\overline{\perp} a} \doteq - \sqrt{ \frac{\widehat{\Lambda}}{3} } Q^{\top}_{ \perp \perp a}.
\end{equation}
Before analysing Eq. (\ref{init2}) let us first focus our attention on (\ref{init1}). If we use the standard expression of the electromagnetic energy-momentum tensor in terms of the electric and magnetic field, $\widecheck{\mathcal{E}}^a$ and $\widecheck{\mathcal{B}}^a$ respectively, then (\ref{init1}) will read
\begin{equation}
    \frac{2}{3} \widecheck{\Lambda} \left( \overline{\mathcal{E}}_a \overline{\mathcal{E}}{}^a + \overline{\mathcal{B}}_a \overline{\mathcal{B}}{}^a \right) + X_{\overline{\perp} \overline{\perp}} \doteq 0
\end{equation}
where $\overline{\mathcal{E}}_a = \omega \widecheck{\mathcal{E}}_a$ and $\overline{\mathcal{B}}_a = \omega \widecheck{\mathcal{B}}_a$ are the unphysical electric and magnetic fields. Hence, this equation is a constraint on the energy density of the electromagnetic field on the conformal boundary.

To see how (\ref{init2}) can be viewed as a constraint on the initial values of $\widecheck{\mathcal{E}}^a$ and $\widecheck{\mathcal{B}}^a $ suppose that the energy-momentum tensor of the previous aeon has the form of electromagnetic field, i.e. 
\begin{equation}
    \widehat{T}_{ab} = 2 \widehat{F}_{ac} \widehat{F}_{b}{}^c - \frac{1}{2} \widehat{g}_{ab} \widehat{F}_{cd} \widehat{F}^{cd}.
\end{equation}
If we introduce its unphysical counterpart in a way that makes the conservation equation $\widehat{\nabla} \widehat{T}_a{}^b =0$ conformally invariant (see Remark \ref{remcons}), i.e.
\begin{equation}
    T_{ab} = \frac{1}{\Omega^2} \widehat{T}_{ab}
\end{equation}
then (\ref{init2}) reads
\begin{equation}
    \frac{2}{3} \widecheck{\Lambda} \overline{T}^{\top}_{\overline{\perp} a} + X^{\top}_{\overline{\perp} a} \doteq  - \frac{1}{3} \widehat{\Lambda} T^{\top}_{\perp a} 
\end{equation}
or
\begin{equation}
    - \frac{4}{3} \widecheck{\Lambda} \left( \overline{\mathcal{E}} \times \overline{\mathcal{B}} \right)_{a} + X^{\top}_{\overline{\perp} a}\doteq   \frac{2 }{3} \widehat{\Lambda} \left( \mathcal{E} \times \mathcal{B} \right)_{a}
\end{equation}
where $\times$ denotes the vector product and $ \mathcal{E}^a = \Omega^{-1} \widehat{\mathcal{E}}{}^a $ and $ \mathcal{B}^a = \Omega^{-1} \widehat{\mathcal{B}}{}^a $ are the unphysical electric and magnetic fields in the previous aeon. Hence, we see that the initial value of the vector product of $\overline{\mathcal{E}}{}^a$ and $\overline{\mathcal{B}}{}^a$ is determined by an analogous quantity computed using the unphysical fields in the previous aeon.

\subsection*{Acknowledgements}
\noindent
JK would like to thank Pawe\l{} Nurowski for suggesting this research problem and pointing out Ref. \cite{eriksson}. The research leading to these results has received funding from the Norwegian Financial Mechanism 2014--2021, project registration number UMO-2019/34/H/ST1/00636.

\section*{Appendix: Electrovacuum spacetimes with regular conformal Bach tensor}

\setcounter{equation}{0}
\renewcommand{\theequation}{A.\arabic{equation}}

Following the discussion at the end of Sec. \ref{bacheq}, we present an example of electrovacuum spacetimes that admit regular conformal Bach equation. This is based on the analysis of the Chevreton tensor given in \cite{eriksson}. As the equations simplify greatly with the use of spinorial formalism, we will employ it here. The signature of the metric will be changed to $(+,-,-,-)$, which is usual in this setting. The other spinorial conventions and notation follow from \cite{javk} with the exception that the complex conjugation will be denoted by $'$.

\subsection*{Bach tensor for source-free electromagnetic fields}

Let $\widecheck{T}_{ab}$ be a source-free electromagnetic energy-momentum tensor. It can be expressed via Maxwell spinor $\widecheck{\phi}_{AB}$ as
\begin{equation}
\widecheck{T}_{ab} = \widecheck{\phi}_{AB} \widecheck{\phi'}_{A'B'}.
\end{equation}
Since 
\begin{equation}
\widecheck{\phi}_{A}{}^B  \widecheck{\phi}_{BC} = \frac{1}{2} \widecheck{\epsilon}_{AB}  \widecheck{\phi}_{CD}  \widecheck{\phi}^{CD},
\end{equation}
then 
\begin{equation}
\widecheck{T}_{a}{}^c \widecheck{T}_{bc} - \frac{1}{4} \widecheck{g}_{ab} \widecheck{T}_{cd} \widecheck{T}^{cd} = 0
\end{equation}
and the Bach tensor reads
\begin{equation}
    \widecheck{B}_{ab} = \frac{1}{2} \widecheck{\Box} \mspace{4mu}  \widecheck{T}_{ab} +  \widecheck{C}_{acbd} \widecheck{T}^{cd}  - \frac{2}{3}  \widecheck{\Lambda} \widecheck{T}_{ab}.
\end{equation}
The Maxwell equations are equivalent to $\widecheck{\nabla}^A{}_{A'} \widecheck{\phi}_{AB} = 0 $, so
\begin{equation}
    0 = \widecheck{\nabla}_{CA'}  \widecheck{\nabla}_A{}^{A'} \widecheck{\phi}^A{}_{B} = - \frac{1}{2} \widecheck{\Box} \mspace{4mu}  \widecheck{\phi}_{CA} + \widecheck{\Box}_{CA} \widecheck{\phi}^A{}_B.
\end{equation}
or with the use of $\widecheck{\Box}_{CA} \widecheck{\phi}^A{}_B = \Psi_{CBAD} \widecheck{\phi}^{AD} +\frac{2}{3} \widecheck{\Lambda}  \widecheck{\phi}_{BC}$,
\begin{equation}
    \widecheck{\Box} \mspace{4mu}  \widecheck{\phi}_{AB} = 2 \Psi_{ABCD} \widecheck{\phi}^{CD} + \frac{4}{3} \widecheck{\Lambda}  \widecheck{\phi}_{AB}. 
\end{equation}
This relation can be used to express $\widecheck{\Box} \mspace{4mu} \widecheck{T}_{ab}$ in terms of lower-order derivatives, i.e.
\begin{equation}
\begin{split}
    \widecheck{\Box} \mspace{4mu} \left( \widecheck{\phi}_{AB} \widecheck{\phi'}_{A'B'} \right) & = 2 \widecheck{\nabla}_{CC'} \widecheck{\phi}_{AB} \widecheck{\nabla}^{CC'} \widecheck{\phi'}_{A'B'}\\
     & - 2 \widecheck{C}_{acbd}\widecheck{T}^{cd} + \frac{8}{3} \widecheck{\Lambda} \widecheck{T}_{ab}.
\end{split}
\end{equation}
Ultimately, the Bach tensor reads
\begin{equation} \label{bachevac}
\widecheck{B}_{ab} = \widecheck{\nabla}_{CC'} \widecheck{\phi}_{AB} \widecheck{\nabla}^{CC'} \widecheck{\phi'}_{A'B'} + \frac{2}{3} \widecheck{\Lambda} \widecheck{\phi}_{AB} \widecheck{\phi'}_{A'B'}.
\end{equation}
or, when written in terms of tensorial quantities (compare with \cite{senerik}),
\begin{equation} \label{bachevact}
    \widecheck{B}_{ab} = \frac{1}{2} \widecheck{g}_{ab} \widecheck{\nabla}_c \widecheck{F}_{de} \widecheck{\nabla}{}^c \widecheck{F}{}^{de} - 2 \widecheck{\nabla}_c \widecheck{F}_{ad} \widecheck{\nabla}{}^c \widecheck{F}_{b}{}^d + \frac{2}{3} \widecheck{\Lambda} \widecheck{T}_{ab}
\end{equation}
where
\begin{equation}
    \widecheck{F}_{ab} = \frac{1}{2} \widecheck{\phi}_{AB} \epsilon_{A'B'} + \frac{1}{2} \widecheck{\phi'}_{A'B'} \epsilon_{AB}
\end{equation}
is the Maxwell tensor.

The first term on the right-hand side of (\ref{bachevac}) will be singular when the equation is written in terms of unphysical tensors associated with $\overline{g}_{ab} = \omega^{-2}\check{g}_{ab}$ and $ \overline{\phi}_{AB} = \omega \widecheck{\phi}_{AB} $ ---which makes the Maxwell equations conformally invariant. Hence, in order for the conformal Bach equation to be regular, an assumption has to be made about the Maxwell spinor. The result in \cite{eriksson} showed that the simple condition $ \widecheck{\nabla}_{CC'} \widecheck{\phi}_{AB} \widecheck{\nabla}^{CC'} \widecheck{\phi'}_{A'B'} = 0$ restricts the spacetime to the Petrov type $\textbf{N}$ or $\textbf{O}$ and vanishing cosmological constant, so in that case the Bach tensor vanishes. 

However, if we consider the following equation
\begin{equation} \label{gbach}
     \widecheck{\nabla}_{CC'} \widecheck{\phi}_{AB} \widecheck{\nabla}^{CC'} \widecheck{\phi'}_{A'B'} = f \widecheck{\phi}_{AB} \widecheck{\phi'}_{A'B'}
\end{equation}
where $f$ is a regular function on the conformal extension $\overline{M}$, we get a nonsingular and nontrivial Bach equation
\begin{equation} \label{bachapp}
    \overline{B}_{ab} = \left(f + \frac{2}{3} \widecheck{\Lambda} \right) \overline{\phi}_{AB} \overline{\phi'}_{A'B'}.
\end{equation}
We see that one possibility for (\ref{gbach}) to be satisfied is
\begin{equation}
    \widecheck{\nabla}_{CC'} \widecheck{\phi}_{AB} = \widecheck{\phi}_{(AB} \widecheck{t}_{C)C'}
\end{equation}
where $\widecheck{t}$ is one of the null tetrad vectors. The other valid approach would be to consider a generalization of condition (24) from \cite{eriksson}, where the homogeneous case was considered ($f=0$). For example,
\begin{equation}
    \widecheck{\nabla}_{CC'} \widecheck{\phi}_{AB} = \eta o_A o_B o_C \iota_{C'}
\end{equation}
where $o_A, \iota_A$ are the spin basis elements and $\eta$ a complex function, could in principle be used to achieve the desired result. However, it can be verified that this assumption leads to a vanishing electromagnetic field if the cosmological constant is not zero. On the other hand, a more general approach with
\begin{equation}
    \widecheck{\nabla}_{CC'} \widecheck{\phi}_{AB} = \eta o_A o_B o_C o_{C'} + \chi \iota_A \iota_B \iota_C \iota_{C'}
\end{equation}
where $\chi$ is another complex function, still leads to the scenario where the null vector $l^a = o^A o^{A'}$ is geodetic and shear-free, but does not correspond to the Weyl spinor of type $N$ or $O$.

It should be stressed that the regularized Bach equation (\ref{bachapp}) is only a necessary condition for an electrovacuum spacetime to fit into the CCC scenario described in Sec. \ref{secconc}. It is not sufficient because of the requirement that the present aeon has a big bang singularity.


\begin{thebibliography}{1}
\bibitem{penrose_ccc}
R. Penrose, \emph{Cycles of Time: An Extraordinary New View of the Universe}, Bodley Head, London, 2010.
\bibitem{penrose_wch}
R. Penrose, \emph{Singularities and Time-Asymmetry}, in S. W. Hawking; W. Israel (eds.). General Relativity: An Einstein Centenary Survey. Cambridge University Press, 581–638, 1979.
\bibitem{cmb1}
D. An, K. Meissner, P. Nurowski, \emph{Ring-type structures in the Planck map of the CMB}, Monthly Notices of the Royal Astronomical Society, \textbf{473} 3, 3251–3255, 2018.
\bibitem{cmb2}
D. An, K. Meissner, P. Nurowski and R. Penrose, \emph{Apparent evidence for Hawking points in the CMB Sky}, Monthly Notices of the Royal Astronomical Society \textbf{495} 3, 3403–3408, 2020.
\bibitem{lopez}
M. Lopez et al., \emph{Searching for ring-like structures in the Cosmic Microwave Background}, in arXiv:2105.03990, 2021.
\bibitem{newman}
E. Newman, \emph{ A fundamental solution to the CCC equations}, Gen Relativ. Gravit. \textbf{46} 1717, 2014.
\bibitem{tod}
K. P. Tod, \emph{The equations of conformal cyclic cosmology}, Gen. Relativ. Gravit. \textbf{47} 17, 2015.
\bibitem{nurek_meissner}
K. Meissner and P. Nurowski, \emph{Conformal transformations and the beginning of the Universe}, Phys. Rev. D \textbf{95} 084016, 2017.
\bibitem{nurowski}
P. Nurowski, \emph{Poincare–Einstein approach to Penrose's conformal cyclic cosmology}, Class. Quantum Grav. \textbf{38} 145004, 2021.
\bibitem{friedrich1}
H. Friedrich, \emph{The asymptotic characteristic initial value problem for Einstein’s vacuum field equations as an initial value problem for a first-order quasilinear symmetric hyperbolic system}, Proc. Roy. Soc. Lond. A, \textbf{378}, 401 1981.
\bibitem{friedrich2}
H. Friedrich, \emph{On the regular and the asymptotic characteristic initial value problem for Einstein’s vacuum field equations}, Proc. Roy. Soc. Lond. A, \textbf{375}, 169, 1981.
\bibitem{anderson}
M. Anderson, \emph{Existence and stability of even-dimensional asymptotically de Sitter spaces}, Annales Henri Poincar\'e, \textbf{6}, 801–820, 2005.
\bibitem{anderson_chrusciel}
M. Anderson and P. Chruściel, \emph{ Asymptotically Simple Solutions of the Vacuum Einstein Equations in Even Dimensions}, Commun. Math. Phys. \textbf{260}, 557–577, 2005.
\bibitem{todqccc}
P. Tod, \emph{Some questions about Conformal Cyclic Cosmology}, in arXiv:2202.10864, 2022.
\bibitem{javk}
J. A. Valiente Kroon, \emph{Conformal Methods in General Relativity}, Cambridge University Press, Cambridge, 2016.

\bibitem{Fri83}
H. Friedrich, \emph{Cauchy problems for the conformal vacuum field equations in General Relativity}, Comm. Math. Phys. \textbf{91}, 445, 1983.

\bibitem{Fri86}
H. Friedrich, \emph{On the existence of n-geodesically complete or future complete solutions of Einstein's field equations with smooth asymptotic structure}, Comm. Math. Phys. \textbf{107}, 587, 1986.

\bibitem{Fri17}
H. Friedrich, Sharp asymptotics for Einstein-$\lambda$-dust flows, Comm. Math. Phys. \textbf{350}, 803, 2017

\bibitem{LueVal13} C. L{\"u}bbe and J.A. Valiente Kroon, \emph{A conformal approach for the analysis of the non-linear stability of pure radiation cosmologies}, Ann. Phys \textbf{328}, 1 (2013).

\bibitem{CarHurVal18} D. Carranza, A. Hursit and J. A. Valiente Kroon. \emph{Conformal wave equations for the Einstein-tracefree matter system}, Gen. Rel. Grav. \textbf{51}, 88, 2019

\bibitem{mannheim}
P. Mannheim, \emph{Making the Case for Conformal Gravity}, Found Phys \textbf{42}, 388–420, 2012.
\bibitem{feff_gr}
C. Fefferman and C. R. Graham, \emph{The Ambient Metric}, Princeton University Press, Princeton, 2012.
\bibitem{kaminski}
W. Kami\'nski, \emph{Well-posedness of the ambient metric equations and stability of even dimensional asymptotically de Sitter spacetimes}, in arXiv:2108.08085, 2021.
\bibitem{schimming}
R. Schimming, \emph{Cauchy’s Problem for Bach’s Equations of General Relativity}, Banach Center Publications 12.1, 225-231, 1984.
\bibitem{bonga}
A. Ashtekar, B. Bonga and A. Kesavan, \emph{Asymptotics with a positive cosmological
constant: I. Basic framework}, Class. Quant. Grav. \textbf{32}, 025004, 2015.
\bibitem{eriksson}
G. Bergqvist and I. Eriksson, \emph{The Chevreton tensor and Einstein-Maxwell spacetimes conformal
to Einstein spaces}, Class. Quant. Grav. \textbf{24}, 3437, 2007.
\bibitem{senerik}
G. Bergqvist, I. Eriksson, and J. M. M. Senovilla, \emph{New electromagnetic conservation laws},
Class. Quant. Grav. \textbf{20}, 2663, 2003.
\end{thebibliography}
\end{document}